  \documentstyle[12pt]{article}
  
\catcode`@=11
\def\secteqno{\@addtoreset{equation}{section}%
\def\theequation{\thesection.\arabic{equation}}}
\catcode`@=12
\secteqno
\newcommand{\be}{\begin{equation}}
\newcommand{\ee}{\end{equation}}
\newcommand{\bref}[1]{(\ref{#1})}
\newcommand{\ep}{\epsilon} \newcommand{\vep}{\varepsilon}

\newcommand{\nn}{\nonumber}
\newcommand{\slp}{/ {\hskip-0.27cm{p}}}

\begin{document}
          \hfill NIIG-DP-01-5

	  \hfill November, 2001
\vskip 20mm

\begin{center} 
{\bf \Large Realization of Global Symmetries\\ in the Wilsonian
 Renormalization Group}

\vskip 10mm
{\large Yuji\ Igarashi, Katsumi\ Itoh and Hiroto\ So$^a$}\par

\medskip
{\it 
Faculty of Education, Niigata University, Niigata 950-2181, Japan\\
$^a$ Department of Physics, Niigata University, Niigata 950-2181, Japan\\
}

\medskip
\date{\today}
\end{center}
\vskip 10mm
\begin{abstract}

We present a method to solve the master equation for the Wilsonian
  action in the antifield formalism. This is based on a representation
  theory for cutoff dependent global symmetries along the Wilsonian
  renormalization group (RG) flow.  For the chiral symmetry, the master
  equation for the free theory yields a continuum version of the
  Ginsparg-Wilson relation.  We construct chiral invariant operators
  describing fermionic self-interactions.  The use of canonically
  transformed variables is shown to simplify the underlying algebraic
  structure of the symmetry.  We also give another non-trivial example,
  a realization of SU(2) vector symmetry. Our formalism may be used
  for a non-perturbative truncation of the Wilsonian action preserving
  global symmetries.

\end{abstract}

\newpage
\setcounter{page}{1}
\setcounter{footnote}{0}
\parskip=7pt

\section{Introduction}

The recent discovery of chiral symmetry on the lattice \cite{Luescher}
has important implications not restricted to lattice theories.  We
strongly believe that it is the prototype of new realization of
symmetry which is not compatible, in the ordinary sense, with a given
regularization.  The symmetry is present, though it undergoes
deformation due to the regularization.

In the Wilsonian renormalization group (RG) \cite{WilsonKogut}, a
promising non-perturbative approach\footnote{For recent progress in
this subject, see, for example, ref. \cite{Roma}.} to continuum
theories, a regulator with the IR cutoff $k$ is introduced to yield the
Wilsonian effective action for lower momentum modes.
The regularization is, often and sometimes inevitably, in conflict
with the standard form of a symmetry. In such a case, the above
realization may be the only possible way of preserving the symmetry.

In previous papers \cite{Igarashi0,Igarashi1,Igarashi2}, we gave
a general formalism for the realization of symmetries, called {\it
renormalized symmetries}, along the RG flow. It applies to global as
well as local symmetries.  What plays a crucial role in describing
renormalized symmetries is the master equation (ME) for the Wilsonian
action.  The ME defines an invariant hypersurface in the theory
space, ie, in the space spanned with couplings.  An important
observation is that once a solution of the ME is found, it stays on
the hypersurface during its RG evolution.  The main task in this
formulation is therefore to solve the ME at some scale of $k$.  In
ref. \cite{Igarashi2}, we gave a general perturbative method for
solving the ME in gauge theories.

It is difficult but highly desirable to construct non-perturbative
solutions to the ME. If such solutions are obtained, they can be used
to make symmetry-preserving truncations of the effective action: the
exact renormalized symmetry is realized in truncated effective action
at every scale along the RG flow.

The purpose of this paper is to discuss, as a first step towards a
consistent truncation, non-perturbative solutions to the ME for global
symmetries. We consider the chiral symmetry and the SU(2) (vector)
symmetry.  Although there are regularizations which manifestly preserve
these symmetries, we take here regularization schemes which are
incompatible with the standard form of symmetries.  By doing this, we
can see how the renormalized symmetries are realized and how their
algebraic structures are related to the ME.

In one of our previous papers \cite{Igarashi0}, we showed that, for
the chiral symmetry, the associated ME and the symmetry
transformations are precisely the continuum counterparts of the
Ginsparg-Wilson (GW) \cite{GW} relation and the L{\"u}scher's chiral
transformations \cite{Luescher}.  In this paper, we use the antifield
formalism of Batalin-Vilkovisky \cite{Batalin} to construct fermionic
interactions which are invariant under the renormalized chiral
transformations. A Wilsonian action consisting of such invariants
solves the ME and gives a symmetry-preserving truncation.

We start our construction of invariant interactions from the average
action for a free theory, which is obtained via the ``block-spin
transformations'' from the UV (microscopic) fields to the IR
(macroscopic) fields.\footnote{These two kinds of fields were introduced
by Ginsparg and Wilson \cite{GW} in lattice theories, and by Wetterich
\cite{Wetterich} in continuum theories.}  After an appropriate canonical
transformation in the field-antifield space, the average action
generates a continuum analog of L{\"u}scher's chiral transformation. It
is also shown that the corresponding algebra is related to the standard
(cutoff independent) chiral algebra via a unitary transformation.  We
may use the algebra to define chiral projections and chiral charges in a
consistent manner.  Based on this representation theory of the cutoff
dependent algebra, we easily obtain invariant interaction terms.

In addition to the chiral symmetry, we also give another non-trivial
example, renormalized non-abelian SU(2) symmetry.  Following the same
procedure used for the chiral symmetry, we find that the ME for the
free theory leads to a simple algebraic GW-like relation for the Dirac
operator.  New generators of the algebra, which depend on the IR
cutoff, can be obtained from the standard generators via a similarity
transformation.  Using this fact, we may obtain a representation of
the renormalized SU(2) symmetry and construct fermionic invariants.
 
This paper is organized as follows. In the next section, we consider
the chiral symmetry using the antifield formalism.  The section 3
describes the application of our formalism to SU(2) symmetry. The final
section is devoted to discussion.

\section{Chiral symmetry }

Let us consider a generic UV action of the Dirac fields,
$S[\bar{\psi},~\psi]$, in d=4 Euclidean momentum space. The UV action
given at a UV scale $\Lambda$ is assumed to be invariant under the
standard chiral transformations:
\begin{eqnarray} 
\delta\psi(p) &=& i~c~\gamma_{5}~ \psi(p),\\ 
\delta \bar{\psi}(p) &=&
i~c~\bar{\psi}(p)~\gamma_5, 
\label{micro-chiraltr} 
\end{eqnarray} 
where $c$ is
a {\it constant} ghost with Grassmann parity $\ep(c)=1$. Let
$\{\psi^{*}(p),~\bar{\psi}^{*}(p)\}$ be the antifields of
$\{\psi(p),~\bar{\psi}(p)\}$. These fields with
$\ep(\psi^{*})=\ep(\bar{\psi}^{*})=0$ play the role of source terms
for $\{\delta \psi,~\delta \bar{\psi}\}$.

According to the general formalism given in ref. \cite{Igarashi1}, we
make the block-spin transformation from the UV fields to the IR fields
$\{\Psi(p),~\bar{\Psi}(p)\}$. This is achieved by using a gaussian term
\begin{eqnarray}
&{}& \int \frac{d^{4}p}{(2\pi)^{4}} \left(\bar{\Psi}(-p)-
f_{k}(p^2)\bar{\psi}(-p)\right)
\alpha^{k}(p^2)  \left(\Psi(p) -  f_{k}(p^2)\psi(p)\right)\nn\\
&~~{}&\equiv 
\int_{p}(\bar{\Psi}-  f_{k}\bar{\psi})(-p)~\alpha^{k}~  (\Psi -  f_{k}\psi)(p),
\label{gauss}
\end{eqnarray}
where $f_{k}(p^2)$ is a function for a ``coarse graining''.  For a
given IR cutoff $k$, $f_{k}(p^2) \approx 0$ for $p^2 < k^2 $, and
$f_{k}(p^2) \approx 1$ for $p^2 > k^2 $.  The cutoff function
$\alpha^{k}(p^2)$ is introduced to relate the UV fields to the IR
fields. It behaves as $f_{k}\alpha^{k} \approx 1$ for $p^2 < k^2$, and
$f_{k}\alpha^{k} \to \infty$ for $p^2 > k^2 $.  At the UV scale
$k=\Lambda$, $f_{\Lambda} \approx 1$, and $\alpha^{\Lambda} \to
\infty$.

Adding the antifield contributions and the gaussian term to 
$S[\bar{\psi},~\psi]$, we have a cutoff dependent action
\begin{eqnarray}
&{}&S_{k}[\psi,\bar{\psi},\Psi,\bar{\Psi}, \psi^*, \bar{\psi}^{*}]
= S_{0}[\psi,~\bar{\psi}]\nn\\
&{}&~~~~ + \int_{p}\left[
\psi^{*}(-p)\delta \psi(p) + 
\delta\bar{\psi}(-p)\bar{\psi}^{*}(p) +  
(\bar{\Psi}-  f_{k} \bar{\psi})(-p)\alpha^{k} 
(\Psi -  f_{k} \psi)(p)\right].
\label{S_k}     
\end{eqnarray}
Using this in the path integral over the UV fields, we may define a
Wilsonian effective action called the average action of the IR fields
\begin{eqnarray}
\exp\left(-W_{k}[\Psi,\bar{\Psi},\Psi^*, \bar{\Psi}^*]/\hbar \right) 
&=& N_{k}
\int {\cal D} \psi {\cal D} \bar{\psi} {\cal D} \psi^{*} {\cal D} \bar{\psi}^{*}\delta\left(f_{k} \Psi^{*}- \psi^{*}\right)
\delta\left(f_{k} \bar{\Psi}^{*}- \bar{\psi}^{*}\right)\nn\\
&{}~~&\times\exp \biggl(
-S_{k}[\psi,\bar{\psi},\Psi,\bar{\Psi}, \psi^*, \bar{\psi}^{*}]/\hbar \biggr),
\label{aveaction}
\end{eqnarray}
where $\Psi^{*}= f_{k}^{-1}\psi$,~and  
$\bar{\Psi}^{*}= f_{k}^{-1}\bar{\psi}$ are
the antifields of $\Psi$ and $\bar{\Psi}$, respectively.  $N_{k}$
is a normalization constant.  Since the regulator $ \alpha^{k}$
behaves as a momentum dependent mass term for the UV fields, the
standard chiral symmetry with transformations \bref{micro-chiraltr} is
broken.  Yet, it is possible to define a
cutoff dependent renormalized symmetry for the average action.  In
order to formulate it, we introduce the antibracket
\begin{eqnarray}
\left(F,~G\right)
&\equiv& \int_{p}\biggl[
\frac{{\partial}^{r} F}{\partial \Psi(-p)} 
\frac{{\partial}^{l} G}{\partial \Psi^{*}(p)}
-\frac{{\partial}^{r} F}{\partial \Psi^{*}(-p)} 
\frac{{\partial}^{l} G}{\partial \Psi(p)}\nn\\
&~~~&+ \frac{{\partial}^{r} F}{\partial \bar{\Psi}(-p)} 
\frac{{\partial}^{l} G}{\partial \bar{\Psi}^{*}(p)}
-\frac{{\partial}^{r} F}{\partial \bar{\Psi}^{*}(-p)} 
\frac{{\partial}^{l} G}{\partial \bar{\Psi}(p)}\biggr]
\label{ab1}
\end{eqnarray}
for any functionals $F$ and $G$ of
$\{\Psi,\bar{\Psi},\Psi^*, \bar{\Psi}^*\}$.
Then, the renormalized symmetry transformations are given by
\begin{eqnarray}
\delta \Psi &=& \left(\Psi,~W_{k}\right)\nn,\\
\delta \bar{\Psi}&=& \left(\bar{\Psi},~W_{k}\right).
\label{macro-chiral}
\end{eqnarray}
Invariance of the action under these transformations is 
expressed by the classical ME
\begin{eqnarray}
(W_{k}, ~W_{k}) =0.
\label{CME}
\end{eqnarray}

For the free theory, it is easy to solve the ME. Note first that in the
path integral \bref{aveaction}, the effective source terms for the UV
fields $\psi$ and $\bar{\psi}$ are proportional to
$(\bar{\Psi}-\Psi^{*}i\gamma_{5}c (\alpha^k)^{-1})$ and $ (\Psi +
i\gamma_{5}c (\alpha^{k})^{-1} \bar{\Psi}^{*})$, respectively.
Therefore, except the bilinear term $\bar{\Psi} \alpha^{k} \Psi$
arising from the block-spin transformation \bref{gauss}, the average
action is a functional of the block-spin variables only through these
combinations.  For instance, its free action takes the form,
\begin{eqnarray}
W_{k}^{(0)}&=&
\int_{p}\Bigl[(\bar{\Psi}-\Psi^{*}i\gamma_{5}c (\alpha^{k})^{-1})(-p)
(D - \alpha^{k})(p) (\Psi + i\gamma_{5}c (\alpha^{k})^{-1} \bar{\Psi}^{*})(p)
\nonumber\\
&+& \bar{\Psi}(-p) \alpha^{k} \Psi(p)\Bigr],
\label{free-average}
\end{eqnarray}
where $D$ is the Dirac operator for the IR fields.
Eqs. \bref{macro-chiral} and \bref{CME} written for the free action
are,
\begin{eqnarray}
\delta \Psi(p) &=& i~c~ \gamma_{5}\left(1- ((\alpha^{k})^{-1}D)(p)\right)
\Psi(p), \nn\\
\delta \bar{\Psi}(-p) &=& i~c~\bar{\Psi}(-p)\left(1- ((\alpha^{k})^{-1}D)(p)\right)\gamma_{5},
\label{chiral-Macro1}
\end{eqnarray}
and 
\begin{eqnarray}
D(p) \gamma_{5} + \gamma_{5} D(p) = 2 ((\alpha^{k})^{-1}D)(p) \gamma_{5} D(p).
\label{GW}
\end{eqnarray}
These are continuum analogs of the L{\"u}sher's chiral
transformations \cite{Luescher} and the GW relation \cite{GW} for
fermion fields on the lattice.

A simple solution to the GW relation is given by
\begin{eqnarray}
D(p)= \frac{\alpha^{k}(p)}{2}\left[1 +\frac{i \slp - \alpha^{k}(p)}
{\sqrt{p^2 +(\alpha^{k})^{2}(p)}}\right].
\label{D}
\end{eqnarray}
This is the continuum version of the lattice solution given by
Neuberger \cite{neu}, and satisfies the boundary condition, $D \to
i\slp /2$ in the limit of $k \to 0$ and $k \to \Lambda$.

We now consider fermionic self-interaction terms. As we explained
already, they must be functions of the block-spin variables,
$(\bar{\Psi}-\Psi^{*}i\gamma_{5}c (\alpha^k)^{-1})$ and $ (\Psi +
i\gamma_{5}c (\alpha^{k})^{-1} \bar{\Psi}^{*})$.  Obviously the
dependence on the antifields makes it to difficult to solve the ME.
This suggests that the original block-spin variables may not suit for
the description of interactions.  Therefore, we look for a canonical
transformation \cite{Batalin1}, \cite{TVV} from $\{\Psi, \bar{\Psi},
\Psi^{*}, \bar{\Psi}^{*}\}$ to a new set of variables $\{\Psi',
\bar{\Psi}', \Psi^{\prime *}, \bar{\Psi}^{\prime *}\}$ in such a way
that the interaction terms can be described only with the new fields
$\{\Psi', \bar{\Psi}'\}$. The canonical transformation would give
us simpler algebraic relations of renormalized symmetry.   For the
free theory under consideration, the generator of such transformation
is found to be
\begin{eqnarray}\hspace{-3mm}
G[\Psi, \bar{\Psi}, \Psi^{\prime *}, \bar{\Psi}^{\prime *}]
  = \int_{p}\left[
  \Psi^{\prime *}(-p) \Psi(p) + \bar{\Psi}(-p) \bar{\Psi}^{\prime *}(p) + 
  \Psi^{\prime *}(-p) c i \gamma_{5} (\alpha^{k})^{-1}(p) \bar{\Psi}^{\prime *}(p)
  \right].
\label{generator}
\end{eqnarray}
This leads to the relations
\begin{eqnarray}
\Psi^{*}(-p) &=& \frac{\partial G}{\partial \Psi(p)} = \Psi^{\prime
*}(-p),\nn\\
\bar{\Psi}^{*}(p)&=& \frac{\partial G}{\partial \bar{\Psi}(-p)}
= \bar{\Psi}^{\prime *}(p),\nn\\
\Psi'(p) &=& \frac{\partial G}{\partial \Psi^{\prime *}(-p)} = \Psi(p) +
 c i \gamma_{5} \alpha^{-1}_{k}(p) \bar{\Psi}^{*}(p),\nn\\
\bar{\Psi}'(-p) &=&\frac{\partial G}{\partial \bar{\Psi}^{\prime *}(p)}
= \bar{\Psi}(-p) + \Psi^{*}(-p) c i \gamma_{5} \alpha^{-1}_{k}(p).
\label{cano-relation}
\end{eqnarray}
Replacing the old variables by new ones in \bref{free-average}, 
we obtain\footnote{Since the jacobian factor from the canonical transformation \bref{cano-relation} is
trivial, there is no additional contribution to the free action.}
\begin{eqnarray}
W_{k}^{(0)}= \int_{p}\left[\bar{\Psi}'(-p) D(p) \Psi'(p) + 
\Psi^{\prime *}(-p) c i \hat{\gamma}_{5}(p) \Psi' - \bar{\Psi}'(-p) c i \gamma_{5}
 \bar{\Psi}^{\prime *}(p)\right],
\label{free-prime}
\end{eqnarray}
where 
\begin{eqnarray}
\hat{\gamma}_{5}(p) &\equiv&  \gamma_{5} \left(1 - 2\hat{D}(p)\right),\nn\\
\hat{D}(p) &\equiv& (\alpha^{k})^{-1}(p)D(p).
\label{gamma-hat-5}
\end{eqnarray}
From now on, we denote the new variables simply as $\{\Psi,~\bar{\Psi}\}$,
discarding primes.  The renormalized chiral transformations are given by
\begin{eqnarray}
\delta \Psi(p) &=&  i~c~ \hat{\gamma}_{5}(p) \Psi(p), \nn\\
\delta \bar{\Psi}(-p) &=& i~c~ \bar{\Psi}(-p)  \gamma_{5}.
\label{chiral-tra-prime}
\end{eqnarray}
The asymmetric transformations on $\Psi$ and $\bar{\Psi}$ appeared in
\bref{chiral-tra-prime} have been known in the lattice chiral theory.
The ME $(W_{k}^{(0)},~W_{k}^{(0)})=0$ yields
\begin{eqnarray}
\hat{D}(p) \hat{\gamma}_{5}(p) + \gamma_{5} \hat{D}(p) =0.
\label{GW2}
\end{eqnarray}
This is another form of the GW relation and extensively used below.  We
find that
\begin{eqnarray}
{\hat{\gamma}_{5}(p)}^{2}=1,~~~~~\hat{\gamma}_{5}^{\dagger}(p)
=\hat{\gamma}_{5}(p),
\label{gamma-hat-5-sq}
\end{eqnarray}
where $D^{\dagger}=\gamma_{5}D\gamma_{5}$. Using the above relations
\bref{GW2} and \bref{gamma-hat-5-sq}, we can define the
L{\"u}sher's chiral projection,\footnote{Note that $ \Psi$ and $\bar{\Psi}$ are
independent and the mass term is $\bar{\Psi}_R {\Psi}_R +
\bar{\Psi}_L {\Psi}_L$ in our notation.}
\begin{eqnarray}
\hat{\Psi}_R &=& \frac{1+\hat{\gamma}_5}{2} \Psi, \nn\\
\hat{\Psi}_L &=& \frac{1-\hat{\gamma}_5}{2} \Psi, \nn\\
\bar{\Psi}_R &=& \bar{\Psi}\frac{1+\gamma_5}{2}, \nn\\
\bar{\Psi}_L &=& \bar{\Psi}\frac{1-\gamma_5}{2}.
\label{chiral-projection}
\end{eqnarray}
The projected fields obey
\begin{eqnarray}
\delta \hat{\Psi}_R &=& i~c~ \hat{\Psi}_R,~~~~~~~
\delta \hat{\Psi}_L = -i~c~ \hat{\Psi}_L, \nn\\
\delta \bar{\Psi}_R &=& i~c~ \bar{\Psi}_R,~~~~~~~
\delta \bar{\Psi}_L = -i~c~\bar{\Psi}_L.
\label{chiral-charge}
\end{eqnarray}
Therefore, we may assign chiral charges $(+,~+,~-,~-)$ to 
$(\hat{\Psi}_R,~\bar{\Psi}_R,~\hat{\Psi}_L,~\bar{\Psi}_L)$. 

It should be remarked that the $\hat{\gamma}_{5}(p)$ is related to
${\gamma}_{5}$ via a unitary transformation
\begin{eqnarray}
\hat{\gamma}_{5}(p) &=& \gamma_{5} \left(1 - 2\hat{D}(p)\right) =
V^{\dagger}(p) \gamma_5 V(p),\nn\\
V(p) &\equiv& \sqrt{1-2{\hat D}(p)},
\label{unitary-tr}
\end{eqnarray}
where $V(p)$ is defined as a power series expansion wrt ${\hat D}(p)$.
The unitarity of $V(p)$ follows from the GW relation \bref{GW}. 
Using $\hat{\gamma}_{\mu}(p) \equiv  V^{\dagger}(p) \gamma_{\mu} V(p)$,
we define operators
\begin{eqnarray}
O_{-}(p) &=& 
\left\{1,~ \hat{\gamma}_{5}(p),~ \frac{i}{2}[\hat{\gamma}_{\mu}(p), 
\hat{\gamma}_{\nu}(p)]\right\} \times (p-{\rm dep.~~factor}),\nn\\ 
O_{+}(p) &=&
\{\hat{\gamma}_{\mu}(p),~\hat{\gamma}_{5}(p)\hat{\gamma}_{\mu}(p)\}
 \times (p-{\rm dep.~~factor}),
\label{0pm}
\end{eqnarray}
which satisfy
\begin{eqnarray}
O_{-}(p)\hat{\gamma}_{5}(p) -\hat{\gamma}_{5}(p)O_{-}=0,~~~~~~~ 
O_{+}(p)\hat{\gamma}_{5}(p) +\hat{\gamma}_{5}(p)O_{+}=0.
\label{Ogamma5}
\end{eqnarray}
It is then easy to list operators which are invariant under the parity
and the chiral transformations \bref{chiral-charge}. 
Using the shorthand notations,
\begin{eqnarray}
&{}&\bar{\Psi}_L O_{-} \hat{\Psi}_R \equiv \int_{p} \bar{\Psi}_{L}(-p)  
 O_{-}(p) \hat{\Psi}_{R}(p),\nn\\
&{}&\biggl(\bar{\Psi}_L O_{-} \hat{\Psi}_L\biggr)\biggl(\bar{\Psi}_R O_{-}
\hat{\Psi}_R\biggr) \equiv  \prod_{i=1}^{4} \int_{p_{i}}\delta 
(\sum_{i=1}^{4} p_{i})\biggl(\bar{\Psi}_{L}(p_{1}) O_{-}(p_{2})
\hat{\Psi}_{L}(p_{2})\biggr)\nn\\
&{}&~~~~~~~~~~~~~ \times\biggl(\bar{\Psi}_{R}(p_{3})
O_{-}(p_{4})\hat{\Psi}_{R}(p_{4})\biggr),
\nn
\end{eqnarray}
we may construct invariants as\\  
\noindent
(1) bilinear operators: $\bar{\Psi}_L O_{\pm} \hat{\Psi}_R  +
 \bar{\Psi}_R O_{\pm} \hat{\Psi}_L$,\\
\noindent
(2) 4-fermi operators:  $\left(\bar{\Psi}_L O_{\pm} \hat{\Psi}_L\right)
\left(\bar{\Psi}_R O_{\pm} \hat{\Psi}_R\right),
~\left(\bar{\Psi}_L O_{\pm} \hat{\Psi}_R\right)
\left(\bar{\Psi}_R O_{\pm} \hat{\Psi}_L\right),\\
 \left(\bar{\Psi}_L
O_{\pm} \hat{\Psi}_R\right)^{2} +\left(\bar{\Psi}_R
O_{\pm} \hat{\Psi}_L\right)^{2}$.

Let us write down some typical invariants. As $O_{\pm}(p)$, we take
those with no additional momentum dependent factors
(cf. \bref{0pm}). The first examples are
\begin{eqnarray}
\biggl(\bar{\Psi}_L \hat{\Psi}_L\biggr)\biggl(\bar{\Psi}_R \hat{\Psi}_R\biggr), 
~~~~~~~~\biggl(\bar{\Psi}_L {\hat \gamma}_{\mu}\hat{\Psi}_R \biggr)^{2}+ 
\biggl(\bar{\Psi}_R {\hat \gamma}_{\mu}\hat{\Psi}_L \biggr)^{2}.
\label{4-fer}
\end{eqnarray}
These are obtained by appropriately replacing the projection operators
in the well-known invariants of the standard chiral symmetry
\begin{eqnarray}
&{}&(\bar{\Psi}_L {\Psi}_L)(\bar{\Psi}_R {\Psi}_R) =\frac{1}{4} \left[(\bar{\Psi}\Psi)^2 - (\bar{\Psi}\gamma_{5}\Psi)^2\right],\nn\\
&{}&(\bar{\Psi}_R {\gamma}_{\mu}{\Psi}_L)^{2}+(\bar{\Psi}_L {\gamma}_{\mu}{\Psi}_R)^{2}=\frac{1}{4}\left[
(\bar{\Psi}\gamma_{\mu}\Psi)^2 + (\bar{\Psi}\gamma_{\mu}\gamma_{5}\Psi)^2\right].
\label{con-4-fer}
\end{eqnarray}
The first invariant in \bref{4-fer} reads
\begin{eqnarray}
\biggl(\bar{\Psi}_L \hat{\Psi}_L\biggr)\biggl(\bar{\Psi}_R \hat{\Psi}_R
\biggr)&=&\frac{1}{4} \biggl[(\bar{\Psi}\Psi)^2 - (\bar{\Psi}\gamma_{5}\Psi)^2
 -2 \left(\bar{\Psi}\hat{D}\Psi\right)(\bar{\Psi}\Psi)\nn\\
&~& +2
\left(\bar{\Psi}\gamma_{5}\hat{D}\Psi\right)
(\bar{\Psi}\gamma_{5}\Psi)+
\left(\bar{\Psi}\hat{D}\Psi\right)^{2}
-\left(\bar{\Psi}\gamma_{5}\hat{D}\Psi\right)^{2}\biggr].
\label{4-fermi-1}
\end{eqnarray} 
A lattice counterpart of this operator \bref{4-fermi-1} expressed by
auxiliary fields was given in ref.  \cite{IN}. In our general
construction, we also have the second invariant in \bref{4-fer} which
cannot be expressed as a polynomial in $\hat{D}$.

The above invariants \bref{4-fer} reduce to the conventional ones
\bref{con-4-fer} in the limit of $\alpha \to \infty~(\hat{D} \to 0)$.
In addition to such invariants, we have another type of invariants which
vanish in this limit.  For example,
\begin{eqnarray}
\biggl(\bar{\Psi}_L \hat{\Psi}_R\biggr)^2 + 
\biggl(\bar{\Psi}_R \hat{\Psi}_L\biggr)^2
= \frac{1}{2}\left[\left(\bar{\Psi}\hat{D}\Psi\right)^2
+\left(\bar{\Psi}\gamma_{5}\hat{D}\Psi\right)^2 \right].
\label{odd-4fer}
\end{eqnarray}
The existence of this kind of invariants is characteristic to the
renormalized chiral symmetry. 

Let $W_{k}^{(1)}[\Psi,~\bar{\Psi}]$ be an action that consists of
invariant operators constructed above.  Since the action $W_{k}^{(1)}$
contains no antifields, the total action
$W_{k}=W_{k}^{(0)}+W_{k}^{(1)}$ is a non-perturbative solution of the
ME \bref{CME}, and gives a symmetry-preserving truncation of the
average action.

\section{Global SU(2) symmetry}

In order to show that our formalism may be applicable to a non-abelian
symmetry, we consider an SU(2) vector symmetry in this section.  The
antifield formalism requires to include constant ghosts $C^{a}$ for
generators $T_{a}= \sigma_{a}/2~~(a=1,2,3)$.  The SU(2)
transformations on the UV fields read
\begin{eqnarray}
\delta \psi(p) &=& i~C^{a}T_{a}~ \psi(p),\nn\\
\delta \bar{\psi}(p) &=& -i~C^{a}~\bar{\psi}(p)~T_{a},\nn\\ 
\delta C^{a} &=& - \frac{1}{2} \vep_{abc} C^{b}C^{c}= - \frac{1}{2}
 (C \times C)^{a}.
\label{micro-SU2tr}
\end{eqnarray}
The free field action of the IR fields is given by
\begin{eqnarray}
W_{k}^{(0)} &=&
\int_{p}\biggl[(\bar{\Psi}-\Psi^{*}iC^{a}T_{a} (\alpha^{k})^{-1})(-p)
(D - \alpha^{k})(p) (\Psi - i (\alpha^{k})^{-1}C^{a}T_{a} \bar{\Psi}^{*})(p)\nn\\
&{}& ~~+ \bar{\Psi}(-p) \alpha^{k} \Psi(p)\biggr] 
- \frac{1}{2} C^{*}_{a}(C \times C)^{a},
\label{free-average-SU2}
\end{eqnarray}
with the matrix $\alpha^{k}$ chosen as
\begin{eqnarray}
\alpha^{k}(p) = \alpha^k_{0}(p) + \alpha^{k}_{3}(p) \sigma_{3}. 
\label{alpha-SU2}
\end{eqnarray}
The regularization using $\alpha^{k}$ clearly violates the standard
SU(2) symmetry.

The action \bref{free-average-SU2} contains bilinear terms of the
antifields. We make a canonical transformation on the IR fields so that
the action becomes linear in the antifields when expressed in terms of
new variables. We take the generator
\begin{eqnarray}
G[\Psi, \bar{\Psi}, \Psi^{\prime *}, \bar{\Psi}^{\prime *}, C, C^{\prime *}]
  &=& \int_{p}\biggl[
  \Psi^{\prime *}(-p) \Psi(p) + \bar{\Psi}(-p) \bar{\Psi}^{\prime *}(p)\nn\\
  &{}& ~-i~\Psi^{\prime *}(-p) (\alpha^{k})^{-1}(p)~C^{a}T_{a}
  \bar{\Psi}^{\prime *}(p)\biggr] + C^{\prime *}_{a}C^{a},
\label{generator-SU2}
\end{eqnarray} 
and find 
\begin{eqnarray}
\Psi'(p) &=& \Psi(p) - i (\alpha^{k})^{-1}(p) C^{a}T_{a}\bar{\Psi}^{*}(p),\nn\\
\bar{\Psi}'(-p) &=&\bar{\Psi}(-p) +i  
\Psi^{*}(-p) (\alpha^{k})^{-1}(p) C^{a}T_{a},
\nn\\
C^{\prime *}_{a} &=& C^{*}_{a} + i \int_{p} \Psi^{*}(-p) 
(\alpha^{k})^{-1}(p) C^{a}T_{a}\bar{\Psi}^{\prime *}(p), 
\label{cano-relation-SU2}\\
\Psi^{*}(-p)&=&\Psi^{\prime *}(-p), ~~~~~ \bar{\Psi}^{*}(p)= \bar{\Psi}^{\prime *}(p).
\nonumber
\end{eqnarray}
In terms of new variables, we obtain
\begin{eqnarray}
W_{k}^{(0)}&=& \int_{p}\biggl[\bar{\Psi}'(-p) D(p) \Psi'(p) + 
\Psi^{\prime *}(-p)  i C^{\prime a}{\hat T}_{a}(p) \Psi' + \bar{\Psi}'(-p) 
 i C^{\prime a} T_{a}
 \bar{\Psi}^{\prime *}(p)\biggr] \nn\\
 &{}& - \frac{1}{2} C^{\prime *}_{a} (C^{\prime} \times C^{\prime})^{a},
\label{free-prime-SU2}
\end{eqnarray}
where the renormalized generators are given by
\begin{eqnarray}
{\hat T}_{a}(p) \equiv T_{a} + [(\alpha^{k})^{-1},~T_{a}]~D(p).
\label{T-hat}
\end{eqnarray}
Then the ME $(W_{k}^{(0)},~W_{k}^{(0)})=0$ yields a GW-like 
algebraic relation
\begin{eqnarray}
D(p)~{\hat T}_{a}(p) - T_{a}~ D(p) =0.
\label{GW-SU2}
\end{eqnarray}
Thanks to this relation, the new generators are shown to satisfy
\begin{eqnarray}
[{\hat T}_{a}(p),~{\hat T}_{b}(p)] = i \vep_{abc} {\hat T}_{c}(p).
\label{SU2-algebra}
\end{eqnarray}
Note that the choice \bref{alpha-SU2} for $\alpha^{k}$ made the diagonal
generator remains invariant, ${\hat T}_{3}(p) = T_{3}$. Then, since
$[T_{3},~D(p)]=0$, the Dirac operator takes the form,
\begin{eqnarray}
D(p) = D_{0}(p) + D_{3}(p) \sigma_{3}.
\label{Dirac-SU2}
\end{eqnarray}
The GW-like relation \bref{GW-SU2} leads to
\begin{eqnarray}
D_{3}(p) = \frac{\alpha^{k}_{3}(p)}{[\alpha^{k}_{0}(p)]^2 - 
[\alpha^{k}_{3}(p)]^2}\left(D_{0}^{2}(p) - D_{3}^{2}(p)\right). 
\label{GW-Dirac}
\end{eqnarray}
The Dirac operator that solve this equation is given by
\begin{eqnarray}
D(p) \equiv \frac{1}{2}\left(\alpha^{k}_{0}(p) + \alpha^{k}_{3}(p)\sigma_{3}
\right)
\left[1 +\frac{i \slp - \alpha^{k}_{0}(p)}
{\sqrt{p^2 +(\alpha^{k}_{0})^{2}(p)}}\right],
\label{Dirac-SU2-2}
\end{eqnarray}
where we impose the boundary condition, $\alpha^{k}_{3}/\alpha^{k}_{0} 
\to 0$ in the limits of $k \to 0$ and $k \to \Lambda$.

Using the relation \bref{GW-Dirac}, one can find the similarity
transformation which relates the standard generators $\{T_{a}\}$ to the
renormalized ones $\{{\hat T}_{a}(p)\}$,
\begin{eqnarray}
{\hat T}_{a}(p) = \exp\left(-\theta (p) T_{3}\right) T_{a} \exp\left
(\theta (p) T_{3}\right), 
\label{sim-tra}
\end{eqnarray}
where 
\begin{eqnarray}
\tanh \theta (p) = \frac{2 \alpha^{k}_{3}(p) D_{0}(p)}
{[\alpha^{k}_{0}(p)]^2 -
[\alpha^{k}_{3}(p)]^2 + 2 \alpha^{k}_{3}(p) D_{3}(p)}.
\label{theta}
\end{eqnarray}
Obviously, the SU(2) algebra in \bref{SU2-algebra} for ${\hat T}_{a}(p)$
is obtained from the standard one by using the similarity
transformation.

The renormalized transformations on the new variables $\{\Psi,~{\bar
\Psi}, C\}$\footnote{From now on, the primes will be discarded.} are
given by
\begin{eqnarray}
\delta \Psi(p) &=& i C^{a}{\hat T}_{a}(p) \Psi(p), \nn\\
\delta {\bar \Psi}(p) &=& -i~C^{a}~\bar{\Psi}(p)~T_{a}, \nn\\
\delta C^{a} &=& - \frac{1}{2}(C \times C)^{a}. 
\label{renor-SU2-tr}
\end{eqnarray}
The presence of the similarity transformations implies that 
bilinear operators 
\begin{eqnarray}
{\bar \Psi}(-p) D(p) \Psi(p),~~~~~~~~
{\bar \Psi}(-p)~\left[\exp\left(\theta(p)T_{3}\right)\right]~\Psi(p),
\label{bilenar}
\end{eqnarray}
are invariant under \bref{renor-SU2-tr}.  Typical 
4-fermi invariant operators are given by
\begin{eqnarray}
\prod_{i=1}^{4} \int_{p_{i}}\delta (\sum_{j=1}^{4} p_{j})~~ {\cal O},
\label{4-fermi-SU2-1}
\end{eqnarray}
where  ${\cal O} ={\cal O}_1$ or ${\cal O}_2$, 
\begin{eqnarray}
\hspace{-5mm}{\cal O}_{1} &=&
\left[{\bar \Psi}(p_1)~\exp\left(\theta(p_2)T_{3}\right)~\Psi(p_2)\right]
\left[{\bar \Psi}(p_3)~\exp\left(\theta(p_4)T_{3}\right)~\Psi(p_4)\right], 
\nn\\
\hspace{-5mm}{\cal O}_{2} &=&
 \left[{\bar \Psi}(p_1)~\exp\left(\theta(p_2)T_{3}\right){\hat T}_{a}(p_2)
\Psi(p_2)\right]
\left[{\bar \Psi}(p_3)~\exp\left(\theta(p_4)T_{3}\right){\hat T}_{a}(p_4)
\Psi(p_4)\right].
\label{4-fermi-SU2-2}
\end{eqnarray}
The average action with such invariants solves the ME and
gives a symmetry-preserving truncation.

\section{Discussion} 

Our method for formulating the renormalized symmetries uses the notion
of the block-spin transformations. The standard form of a symmetry is
assumed to be realized in the UV action. The block-spin variables are
useful in deriving the ME, which ensures the presence of the
renormalized symmetry. In solving the ME, however, the variables are
found not to be convenient due to the following reasons. (1) The
antifields appear in the interaction terms and therefore the symmetry
transformations will depend on the interaction terms. (2) As seen in the
case of SU(2) symmetry, the average action generally becomes nonlinear
in the antifields even for a free theory.
 
The above problems are closely related to the choices of the dynamical
variables in the average action. We have shown that the use of new
variables obtained by the canonical transformations makes the average
action linear in antifields, and simplifies representations of the
underlying algebra of the renormalized symmetries. For the SU(2)
symmetry, we have obtained the closed algebra. The GW relations in both
symmetries obtained from the ME for the free actions played an important
role in constructing invariant operators.

Our discussion for the fermionic systems may be extended to SU(N) $(N
> 2)$ symmetry in such a way that the Cartan's subalgebra remains
unchanged. In our specific examples including these cases, the
renormalized symmetries are realized in an asymmetric way: the
transformations on $\Psi$ undergo deformation, while those on
$\bar{\Psi}$ remain intact. 

The solutions we have constructed in this paper are those to the
classical ME rather to the quantum ME. Let $W_{k}^{(q)}$ be a solution
of the quantum ME: $(W_{k}^{(q)}, W_{k}^{(q)})/2 - \hbar \Delta
W_{k}^{(q)}=0$. The $\Delta$-derivative term may arise from the jacobian
factor \cite{TVV} of the functional measure associated with the
renormalized symmetry transformations. For an anomaly-free theory, all
terms arising from the jacobian factor must be the coboundary terms
which can be removed by introducing a suitable counter action ${\tilde
W}$, ie, $( W_{k}^{(q)},~{\tilde W}) - \hbar \Delta W_{k}^{(q)} =0$.
Furthermore, if the counter action ${\tilde W}$ has no antifield
dependence, we find that $( W_{k}^{(q)} -{\tilde W}, W_{k}^{(q)} -
{\tilde W} )=0 $.  Therefore, the action $W_{k}^{(c)} \equiv W_{k}^{(q)}
- {\tilde W}$ obeys the classical ME.  Note that the concrete form of
the counter action ${\tilde W}$ depends on the UV regularization scheme
which is needed to make the $\Delta$-derivative well-defined. Up to this
action, the $W_{k}^{(q)}$ is equivalent to $W_{k}^{(c)}$. Thus, we may
conclude that our solutions of the classical ME provide a reasonable
truncation for the average action.

It is worth studying the evolution of couplings for the invariant
operators.  This may be achieved by solving the exact flow equations.
We expect that these considerations provide us with some important clues
for the realization of gauge (BRS) symmetries along the RG flow.

\section*{Acknowledgments} 

This work is supported in part by the Grants-in-Aid for Scientific
Research No. 12640258, 12640259, and 13135209 from the Japan Society for
the Promotion of Science.


\end{document}